\documentclass[reprint,letterpaper,twocolumn,amssymb,nobibnotes,showpacs, aps, pra]{revtex4-1}
\usepackage{graphicx}

\begin{document}

\title{Two-Color Magneto-Optical Trap with Small Magnetic Field for Ytterbium}

\author{Akio Kawasaki}
\email{akiok@mit.edu}
\affiliation{Department of Physics, MIT-Harvard Center for Ultracold Atoms and Research Laboratory of Electronics, Massachusetts Institute of Technology, Cambridge, Massachusetts 02139, USA}

\author{Boris Braverman}
\affiliation{Department of Physics, MIT-Harvard Center for Ultracold Atoms and Research Laboratory of Electronics, Massachusetts Institute of Technology, Cambridge, Massachusetts 02139, USA}

\author{QinQin Yu}
\affiliation{Department of Physics, MIT-Harvard Center for Ultracold Atoms and Research Laboratory of Electronics, Massachusetts Institute of Technology, Cambridge, Massachusetts 02139, USA}

\author{Vladan Vuleti${\rm{\acute{c}}}$ }
\affiliation{Department of Physics, MIT-Harvard Center for Ultracold Atoms and Research Laboratory of Electronics, Massachusetts Institute of Technology, Cambridge, Massachusetts 02139, USA}

\date{\today}

\begin{abstract}
We report a two-color magneto-optical trap (MOT) for ytterbium atoms operating at a low magnetic field gradient down to 2 G/cm where a conventional MOT using the singlet transition ($\mathrm{6s}^2 \; ^1\mathrm{S}_0\rightarrow \mathrm{6s6p} \; ^1\mathrm{P}_1$) is unable to trap atoms. By simultaneously applying laser light on both the broad-linewidth singlet transition and the narrow-linewidth triplet transition ($\mathrm{6s}^2 \; ^1\mathrm{S}_0\rightarrow \mathrm{6s6p} \; ^3\mathrm{P}_1$), we load and trap $4.0 \times 10^5$ atoms directly from an atomic beam at 700 K.  In the two-color MOT, the slowing and trapping functions are separately performed by the singlet transition light and the triplet transition light, respectively.  The two-color MOT is highly robust against laser power imbalance even at very low magnetic field gradients.  
\end{abstract}

\pacs{37.10De, 37.10Gh, 37.10Vz}
\maketitle
\section{Introduction}\label{Introduction}
The magneto-optical trap (MOT) \cite{PhysRevLett.59.2631} has become the standard way of laser cooling and trapping neutral atoms. Due to its simplicity and robustness, the MOT has enabled many experiments with ultracold neutral atoms, such as quantum gases \cite{Science.269.198,PhysRevLett.75.3969,Science.285.1703}, cavity QED systems \cite{JPhysB.38.s551}, and atomic clocks \cite{1407.3493}.  In a MOT, light slightly red-detuned from an atomic cycling transition provides cooling and trapping forces through the Doppler and Zeeman effects, respectively. In a conventional MOT for alkali atoms \cite{PhysRevLett.59.2631}, both functions are performed by 
a single laser.  However, the cooling and trapping can in principle be performed by light fields addressing two different transitions. At the cost of some technical complexity, the additional available parameters, such as transition linewidth, and laser intensity and detuning, may allow one to optimize the cooling and trapping functions separately, and improve the performance over the standard one-color MOT. 

One class of atoms with two transitions of different linewidth are the alkaline earth and alkaline earth-like atoms. With two electrons in the outermost shell, these atoms have a broad-linewidth singlet transition and a narrow-linewidth triplet transition that are both suitable for a MOT. The broad singlet transition is excellent for slowing fast atoms, enabling their loading into a trap of finite depth. However, the singlet MOT has a minimum achievable Doppler temperature of $\hbar \Gamma_s/(2k_B) \sim 1$ mK, set by the large transition linewidth of $\Gamma_s/(2\pi)\simeq30$ MHz. On the other hand, it requires a large magnetic field gradient of typically $B'=50$ G/cm \cite{PhysRevA.68.055401,PhysRevA.61.051401,PhysRevA.59.R934} to match the Zeeman shift to $\Gamma_s$ over a characteristic distance of 1 cm. The narrow triplet transition allows one to cool atoms down to temperatures of a few $\mu$K, corresponding to transition linewidths in the range of $\Gamma_t/(2\pi)=(1-100)$ kHz, but the atoms have to be initially slow to be captured into a triplet MOT because the slowing force associated with the small scattering rate $\le\Gamma_t$ is too weak to capture fast atoms.  Alkaline earth (-like) atoms have been cooled down to $\mu$K temperatures by applying light on the two transitions sequentially \cite{PhysRevA.68.011403,ApplPhysExp.2.072501,Metrologia.50.119,PhysRevLett.103.063001,PhysRevA.82.043429,PhysRevLett.82.1116,PhysRevA.60.R745,PhysRevLett.106.153201,ChinPhysLett.26.083702,RevSciInstrum.84.043109}.
  
\begin{figure}[!b]
	\begin{center}
 \includegraphics[width=1.0\columnwidth]{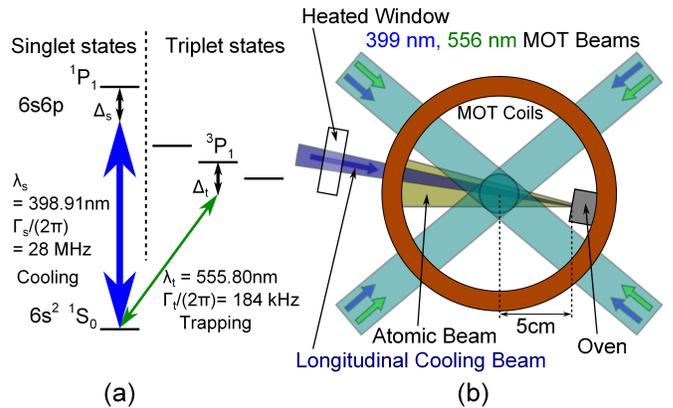}
  \caption{
(a) Energy level diagram for ytterbium. We use the broad singlet transition (blue arrow) and narrow triplet transition (green arrow) for the TCMOT.
(b) Schematic view of the experimental apparatus.} 
  \label{Apparatus_and_Levels}
 \end{center}
\end{figure}

Among the alkaline earth-like atoms, ytterbium is widely used for the study of quantum gases \cite{PhysRevLett.91.040404,PhysRevLett.98.030401}, atomic clocks \cite{Science.341.1215}, and quantum measurements \cite{PhysRevLett.102.033601}.  As shown in Fig. \ref{Apparatus_and_Levels} (a), the $\mathrm{6s}^2 \; ^1\mathrm{S}_0\rightarrow \mathrm{6s6p} \; ^1\mathrm{P}_1$ transition (singlet transition) has a wavelength of $\lambda_s=399$ nm and a linewidth of $\Gamma_{{\rm s}}/(2\pi)=28$ MHz. The $\mathrm{6s}^2 \; ^1\mathrm{S}_0\rightarrow \mathrm{6s6p} \; ^3\mathrm{P}_1$ transition (triplet transition) has a wavelength of $\lambda_t=556$ nm and a linewidth of $\Gamma_{{\rm t}}/(2\pi)=184$ kHz. 

Here, we report the realization of a two-color MOT (TCMOT) for ytterbium where 399 nm light and 556 nm light are simultaneously applied. By separating the functions of cooling and trapping, the TCMOT can operate at a small magnetic field gradient $B'$ where neither light on its own can cool and trap atoms efficiently.  Our setup requires only $B'=5$ G/cm to trap a number of atoms similar to a conventional MOT operating on the singlet transition (singlet MOT) that uses a ten times larger field gradient $B'$.  Placing the atom source close to the MOT region enables us to operate the TCMOT without a Zeeman slower.  

The TCMOT is most useful when electric power is limited, such as in portable or space-borne systems. 
In our case, the electric power necessary for the TCMOT was 40 times smaller than that for the singlet MOT with $B'=45$ G/cm.  The electric power consumption in a 2D singlet MOT could also be substantially reduced with a 2D TCMOT. Furthermore, the TCMOT could be useful in situations where a small magnetic field is required to reduce the magnetization of materials around the atoms, e.g. in high-performance optical-transition clocks \cite{Science.341.1215}. 

\begin{figure}[!t]
	\begin{center}
		\includegraphics[width=0.95\columnwidth,clip]{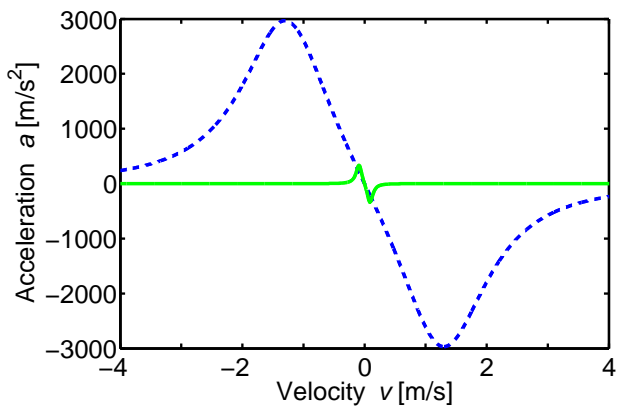}
		\caption{Calculated velocity dependence of the light-induced acceleration due to the singlet (blue dashed line) and triplet (green solid line) transitions. Parameters were set as $\Delta_{{\rm s}}=-0.7\Gamma_{\rm s}$, $\Delta_{{\rm t}}=-5.5\Gamma_{\rm t}$, $I_{{\rm s}}=0.043I_{{\rm sat,s}}$, and $I_{{\rm t}}=27I_{{\rm sat,t}}$.
		} 
		\label{PotVSGDetTheor}
	\end{center}
\end{figure}

Experiments requiring the trapping and cooling of ytterbium down to $\sim5 \; \mu$K in a MOT operated on the triplet transition (triplet MOT) have so far used one of the following three methods. The first method is to load the singlet MOT to trap a large number of atoms, and then transfer the atoms into a triplet MOT to cool them further \cite{PhysRevA.68.011403,ApplPhysExp.2.072501,Metrologia.50.119,PhysRevLett.103.063001,PhysRevA.82.043429}. This two-stage MOT is commonly used for other elements as well \cite{PhysRevLett.82.1116,PhysRevA.84.061406,PhysRevA.84.063420,PhysRevLett.107.190401,PhysRevA.85.051401,QuantumElectron.44.515}.  The second method is to use a Zeeman slower to decelerate an atomic beam, allowing the atoms to be loaded directly into the triplet MOT \cite{PhysRevA.60.R745,PhysRevLett.106.153201,ChinPhysLett.26.083702}. The third method is to use a 2D singlet MOT to load a 3D triplet MOT \cite{RevSciInstrum.84.043109}.  All of these methods require high magnetic fields, because the singlet MOTs (both 2D and 3D) typically requires magnetic field gradients in the range of 30-50 G/cm, while a Zeeman slower requires a magnetic field on the order of $100$ G. 

Methods similar to ours, where 399 nm and 556 nm lights are applied simultaneously, have been very recently reported. Ref. \cite{JPhysSocJpn.83.014301} describes the cooling of ytterbium atoms in an optical lattice by optical molasses, rather than a MOT. During the preparation of this manuscript, another scheme for a small field gradient MOT was reported, where 556 nm light is sent in the central region, surrounded by a 399 nm light shell without spatial overlap between the two light fields. The optimal magnetic field gradient in that experiment was similar to ours, $B'\simeq5$ G/cm, but the 556 nm light was spectrally broadened, and a Zeeman slower was used to increase the flux of slow atoms \cite{1412.2854}.

\section{Principle of the TCMOT}\label{Model}

We can understand the TCMOT as an atom trap where the cooling and trapping functions are separately performed by two light fields.  Each transition's contribution to the force on an atom in the trap at position $z$ moving with velocity $v$ is modeled simply as the sum of the forces $F_\pm$ from counter-propagating beams with intensities $I_\pm(z)$ \cite{PhysRevA.70.063413}:

\begin{eqnarray}
F_\pm (v,z)  =\pm \frac{\hbar k \Gamma}{2}  \frac{I_\pm (z)/I_{{\rm sat}}}{1+I_{\pm} (z)/I_{{\rm sat}}+4\left(\frac{\Delta \pm kv\mp\mu B' z}{\Gamma}\right)^2}
\end{eqnarray}
with wavenumber $k=2\pi/\lambda$, transition linewidth $\Gamma$, transition saturation intensity $I_{{\rm sat}}$, detuning of the light from resonance $\Delta$, and magnetic moment associated with the transition $\mu$.

\begin{figure}
 \begin{center}
 \includegraphics[width=0.95\columnwidth,clip]{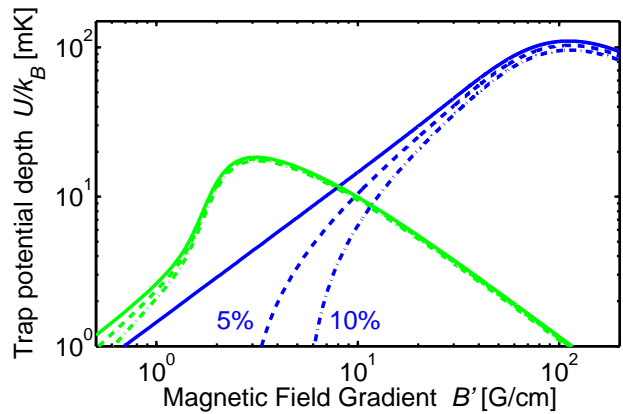}
  \caption{Calculated trapping potential depth for atoms at rest in the singlet MOT (blue lines) and the triplet MOT (green lines) versus $B'$: the solid lines, dashed lines, and dash-dotted lines are for an ideal MOT, and power imbalances of 5\%, and 10\% between the counter-propagating beams, respectively. 
Parameters were set as $\Delta_{{\rm s}}=-0.7\Gamma_{\rm s}$, $\Delta_{{\rm t}}=-5.5\Gamma_{\rm t}$, $I_{{\rm s}}=0.043I_{{\rm sat,s}}$, $I_{{\rm t}}=27I_{{\rm sat,t}}$, and beam size 1 cm.} 
  \label{PotVSFieldTheor}
 \end{center}
\end{figure}

Figure \ref{PotVSGDetTheor} shows the calculated slowing force for each of the transitions, and Fig. \ref{PotVSFieldTheor} shows the calculated trap potential depth for an atom at rest. According to Fig. \ref{PotVSGDetTheor}, the broad singlet transition can address a substantially larger velocity range, and thus has a significantly larger contribution to slowing the atoms regardless of the field gradient $B'$. Note that the integrated flux in an atomic beam up to a capture velocity $v_{\rm c}$ scales as $v_{\rm c}^4$.  This means that for the conditions displayed in Fig. \ref{PotVSGDetTheor}, the loading flux is $5000$ times larger for the singlet light than for the triplet light.  On the other hand, from Fig. \ref{PotVSFieldTheor}, we see that at $B'\lesssim 7$ G/cm, the trapping potential in the triplet MOT is deeper than in the singlet MOT, as the narrower triplet transition provides more frequency discrimination and thus a stronger magnetic-field-dependent restoring force. For the same reason, polarization impurities and intensity imbalance between two counter-propagating MOT beams has a larger effect on the singlet MOT than on the triplet MOT. As the intensity imbalance increases, the field gradient $B'$ necessary to attain a certain trapping depth increases. We note that even for carefully balanced beams, reflections from various surfaces can easily give rise to interference that produces intensity variation of 5-10\% across the beam profile.  

By applying both the singlet transition and triplet transition light simultaneously, it is possible to produce a TCMOT that combines the large cooling force of the singlet transition with the deep trapping potential of the triplet transition at small field gradient $B'$.  This is the main idea of our approach.  

\section{Experimental Setup}

Our apparatus is shown in Fig. \ref{Apparatus_and_Levels} (b). All trapping experiments were performed with $^{174}$Yb atoms, the isotope with the largest natural abundance of 32\%. The frequency detunings of the lasers are measured from the $^{174}$Yb atomic resonances. Typically, the detunings for 399 nm and 556 nm light are $\Delta_{{\rm s}}=-0.7\Gamma_{\rm s}$ and $\Delta_{{\rm t}}=-3.7\Gamma_{\rm t}$, respectively. The MOT beams have a $1/e^2$ diameter of 1 cm.  The total beam intensities for the 399 nm and 556 nm MOT beams are $I_{{\rm s}}=0.26 I_{{\rm sat,s}}$ and $I_{{\rm t}}=160I_{{\rm sat,t}}$, respectively, where $I_{{\rm sat,s}}=58$ mW/cm$^2$ and $I_{{\rm sat,t}}=0.138$ mW/cm$^2$ are the corresponding saturation intensities. 

The atomic source is an oven heated up to 700 K, located 5 cm from the center of the MOT. In addition to the six MOT beams, we use a beam of 399 nm light, counterpropagating with respect to the atomic beam, in order to slow down a portion of the atomic beam.  On the path of the slowing beam inside the vacuum chamber, a heated window at a temperature of 600 K with a power consumption of 12 W is placed in order to prevent the atomic beam from coating the vacuum chamber viewport. The slowing beam with a power of $P_{\rm CP}=3$ mW is detuned by $\Delta_{\rm CP}=-4.6\Gamma_{\rm s}$. This beam is focused over a 65 cm distance from a 1.3 cm initial diameter to a tightest waist of 13 $\mu$m at the collimating hole for the atomic beam. 

The 399 nm light for the MOT and the slowing beam are produced by an external cavity diode laser system locked to the atomic transition by dichroic atomic vapor laser lock \cite{ApplOpt.37.3295,OptLett.28.245}, resulting in a laser linewidth of $\sim4$ MHz. The 556 nm light is produced by a frequency-doubled fiber laser that is locked to a reference cavity, with a short-term linewidth of 35 kHz.

We use two CCD cameras for atom number and atom cloud RMS radius measurements, and an avalanche photodiode for loading-time measurements.  The typical measurement sequence consists of loading atoms into the TCMOT, taking an image of the TCMOT, switching off the 399 nm light, and taking an image of the triplet MOT. Five measurements were averaged for each data point.

\section{Experiments and Interpretation}\label{Result}
\subsection{Properties of the TCMOT}
\begin{figure}[!tb]
 \begin{center}
 \includegraphics[width=1.0\columnwidth,clip]{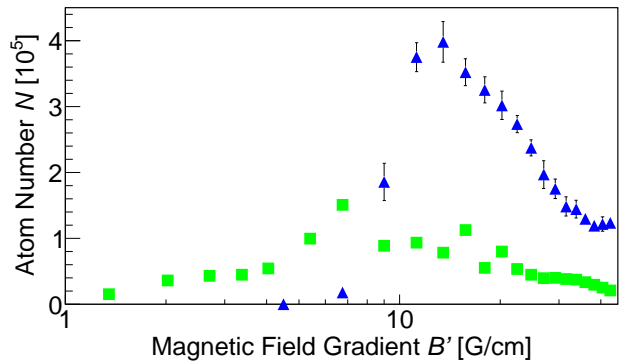}
  \caption{Comparison of the atom number in the singlet MOT (blue triangles) and the TCMOT (green squares) as a function of quadrupole field gradient $B'$: for $B' \leq 6$ G/cm, the singlet MOT does not trap any atoms. Other parameters were $\Delta_{{\rm s}}=-0.7\Gamma_{\rm s}$, $\Delta_{{\rm t}}=-3.7\Gamma_{\rm t}$, $I_{{\rm s}}=0.26I_{{\rm sat,s}}$, and $I_{{\rm t}}=160I_{{\rm sat,t}}$. } 
  \label{BH}
 \end{center}
\end{figure}

\begin{figure*}[!t]
 \begin{center}
 \includegraphics[width=1.7\columnwidth,clip]{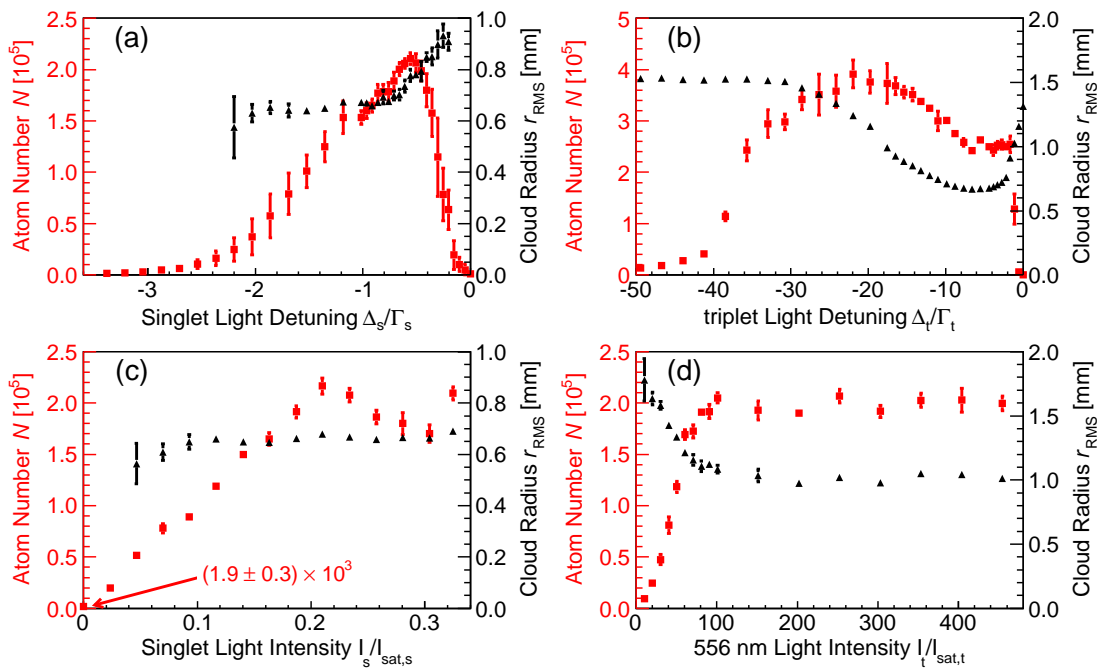}
  \caption{Characterization of the TCMOT: atom number $N$ (red squares) and RMS cloud size $r_{\rm RMS }$ (black triangles) are plotted against (a) $\Delta_{{\rm s}}$, (b)  $\Delta_{{\rm t}}$, (c) $I_{{\rm s}}$, and (d)$I_{{\rm t}}$. We varied only one parameter for each graph, shown on the horizontal axis. Fixed parameters were $\Delta_{{\rm s}}=-0.7\Gamma_{\rm s}$, $\Delta_{{\rm t}}=-3.7\Gamma_{\rm t}$, $I_{{\rm s}}=0.26I_{{\rm sat,s}}$, $I_{{\rm t}}=160I_{{\rm sat,t}}$, and $B'=6.75$ G/cm.} 
  \label{Hybrid}
 \end{center}
\end{figure*}

We first compared the TCMOT with the singlet MOT. Fig. \ref{BH} shows the dependence of atom number $N$ on magnetic field gradient $B'$ for the singlet MOT and the TCMOT. Our singlet MOT traps a maximum $N=4.0 \times 10^5$ at a relatively low $B'$ of 13.5 G/cm, compared to a typical $B'$ for a singlet MOT of 30-45 G/cm \cite{PhysRevA.68.055401,PhysRevA.61.051401}. This is likely due to careful adjustment of the retroreflection and the beam intensity balance, as well as the purity of the circular polarization. However, below 13.5 G/cm, $N$ quickly decreases to zero, and the singlet MOT vanishes altogether at 6 G/cm. This is consistent with a residual 5-10\% beam intensity imbalance displayed in Fig. \ref{PotVSFieldTheor}.  On the other hand, the TCMOT still had $N=1.5 \times 10^5$ at 6.75 G/cm, comparable to the largest $N$ for our singlet MOT. 
The somewhat smaller peak atom number for the TCMOT than for the singlet MOT is qualitatively explained by strong saturation of the triplet transition that results in a reduced scattering rate and slowing force on the singlet transition. The RMS radius of the singlet MOT is $r_{\rm RMS}=2.1$ mm, compared to $r_{\rm RMS}=0.7$ mm for the TCMOT. This factor of 3 in $r_{\rm RMS}$, together with a factor of 8 in $N$, results in a $\sim200$ times larger atom density in the TCMOT at 6.75 G/cm. 

Figure \ref{Hybrid} shows the atom number $N$ and cloud size $r_{\rm RMS}$ of the TCMOT as a function of the detuning $\Delta_{{\rm s}}$, $\Delta_{{\rm t}}$ and the intensity $I_{{\rm s}}$, $I_{{\rm t}}$ of two lasers. Fig. \ref{Hybrid} (a) shows the behavior of the TCMOT versus $\Delta_{{\rm s}}$. The largest $N$ is observed around $\Delta_{{\rm s}}=-0.7\Gamma_{\rm s}$. The TCMOT traps more atoms at large detuning $|\Delta_{{\rm s}}|$ compared to the singlet MOT.  This is likely because for the singlet MOT the restoring force becomes too small at large detuning, while for the TCMOT the restoring force is provided by the 556 nm laser.  Fig. \ref{Hybrid} (b) shows the dependence of the atom number $N$ on $\Delta_{{\rm t}}$. The sharp increase of $N$ near the resonance happens over a frequency range comparable to $\Gamma_{{\rm t}}$, while the largest atom number is observed at $\Delta_{{\rm t}}=-22\Gamma_{\rm t}$. As a function of beam intensity, the atom number increases with $I_{{\rm s}}$ or $I_{{\rm t}}$, until a plateau is observed at $I_{{\rm s}}\gtrsim 0.2I_{{\rm sat,s}}$, and $I_{{\rm t}}\gtrsim 100I_{{\rm sat,t}}$, as shown in Fig. \ref{Hybrid} (c) and (d). 

When $I_{{\rm s}}=0$, we observe a pure triplet MOT with $N=1.9 \times 10^3$, loaded directly from a thermal atomic beam (see Fig. \ref{Hybrid} (c)). With optimized parameters of $\Delta_{{\rm t}}=-14.7\Gamma_{\rm t}$, $\Delta_{\rm CP}=-7.7\Gamma_{\rm s}$, and $P_{\rm CP}=14$ mW, the atom number increased to $(1.2 \pm 0.1) \times 10^4$. Although this is four orders of magnitude smaller than the atom number reported in Ref. \cite{PhysRevA.60.R745}, to the best of our knowledge, this is the first observation of a triplet MOT being loaded from an atomic beam without a Zeeman slower or a 2D MOT.

For the TCMOT, with the optimized parameters $B'=6.75$ G/cm, $\Delta_{{\rm t}}=-22\Gamma_{\rm t}$, $\Delta_{{\rm s}}=-0.7\Gamma_{\rm s}$, $I_{{\rm t}}>100$ $I_{{\rm sat,t}}$, and $I_{{\rm s}}>0.20$ $I_{{\rm sat,s}}$, we observe the largest atom number $N=4.0 \times 10^5$. This is more than two orders of magnitude smaller than the largest $N$ reported in an ytterbium MOT \cite{PhysRevA.60.R745,1412.2854}, primarily because our apparatus without Zeeman slower has been designed for cavity QED experiments that do not require a large number of atoms. Together with the oven temperature and the theoretically calculated capture velocity for the TCMOT of 8 m/s, the measured atomic beam flux $2.2 \times 10^{10}$/s gives an estimated loading rate of $6 \times 10^5$/s, close to the measured value of $2 \times 10^5$/s. Typical values for the atom flux, loading rate, and $N$ with a Zeeman slower are $10^{10}$ s$^{-1}$, $10^{7}$ s$^{-1}$, and $10^{7}$ for the singlet MOT \cite{PhysRevA.85.035401,PhysRevA.61.051401}, and therefore with a Zeeman slower, we would expect the TCMOT to perform at least as well as a conventional singlet MOT. 

The temperature for the TCMOT is 1 mK, slightly higher than the singlet transition Doppler limit of $\hbar \Gamma_s/(2k_B)=0.67$ $\mu$K.  This is 25 times larger than the expected temperature of $\hbar \Delta_{\rm t}/(2k_B)=66$ $\mu$K at $\Delta_{{\rm t}}=-3.7\Gamma_{\rm t}$ \cite{JOptSocAmB.6.2084} for the triplet MOT at the given detuning.  This implies that the 556 nm light only traps the atoms but does not cool them down substantially in the presence of the scattering and the line broadening by the 399 light that heats the atoms toward the Doppler limit of the singlet transition.

\subsection{Experiments testing the model}

To show the TCMOT relies on the simultaneous but independent effects from the singlet and the triplet light fields, we verified that when the two light fields are alternated sufficiently fast, the atom number is similar to that of the TCMOT at half the light intensity (Table \ref{TheoryTest}), while each laser alone traps a much smaller number of atoms.  

To verify that the 556 nm light provides the dominant contribution to the trap confinement, we blocked the center of the 399 nm light with a 5.7 mm diameter disk (blocked TCMOT).  We observe $r_{\rm RMS}=0.67$ mm for the blocked TCMOT, quite close to $r_{\rm RMS}$ for the triplet MOT (Table \ref{TheoryTest}).  This value, also close to $r_{\rm RMS}= 0.69$ mm observed for the TCMOT, shows that the singlet transition does not contribute significantly to the trapping.  This method is similar to the core-shell MOT described in Ref. \cite{1412.2854}, and the result is consistent with Ref. \cite{1412.2854} in that adding 399 nm light to the triplet MOT enhances the atom number.

\begin{table}[!t]
 \caption{$N$ and $r_{\rm RMS} $  in the MOT in different conditions. Parameters are fixed at $\Delta_{{\rm s}}=-0.7\Gamma_{\rm s}$, $\Delta_{{\rm t}}=-3.7\Gamma_{\rm t}$, $I_{{\rm s}}=0.26I_{{\rm sat,s}}$, $I_{{\rm t}}=160I_{{\rm sat,t}}$, and $B'=6.75$ G/cm. The errors shown are statistical errors; systematic errors are estimated to be 10\%.} 
\begin{tabular}{lcrclcrcl}
  \hline   \hline 
Condition	&\hspace{2mm} &\multicolumn{3}{c}{$N$ ($\times 10^3$)} &\hspace{2mm} & \multicolumn{3}{c}{$r_{\rm RMS} $ [mm]}\\
  \hline   
Full power TCMOT 						& &	$243$	& $\pm$ & $10$	& & 0.69	& $\pm$ & 0.01	\\ 
1/2 power TCMOT 					& &	$60$	& $\pm$ & $8$	& & 0.69	& $\pm$ & 0.01	\\ 
25 kHz singlet/triplet switching 	& &	$25$	& $\pm$ & $3$	& & 0.80	& $\pm$ & 0.09	\\ 
Pure singlet MOT 					& &	$5.4$	& $\pm$ & $0.7$	& & 2.04	& $\pm$ & 0.03	\\ 
Pure triplet MOT 					& &	$1.1$	& $\pm$ & $0.2$	& & 0.61	& $\pm$ & 0.02	\\ 
Blocked TCMOT 						& &	$16$	& $\pm$ & $1 $	& & 0.67	& $\pm$ & 0.01	\\ 
  \hline   \hline 
\end{tabular}
\label{TheoryTest}
\end{table}

\subsection{Two-color MOT for other atomic species}
The model described in Section \ref{Model} predicts the range of ratios between the linewidths of the two transitions where the TCMOT is expected to work. The TCMOT requires that atoms cooled by the broad transition are cold enough to be confined by the narrow transition. The lower bound of the narrow transition linewidth for the TCMOT is $k_B T_s\simeq U_t$, where $k_B T_s\sim\hbar \Gamma_s$ is the Doppler temperature of the singlet MOT. The trap depth scales as $U_t\sim \Gamma_t^{3/2}$, as the force scales as $\Gamma_t$ and the spatial range over which the transition is near resonant scales as $\Gamma_t^{1/2}$ because the saturation-broadened linewidth scales as $\Gamma_t^{1/2}$, assuming a constant laser intensity. This results in a lower bound of $\Gamma_t/(2\pi))$ of 35 kHz for Yb, implying a maximum linewidth ratio of $\Gamma_{\rm s}/\Gamma_{\rm t}\sim800$. 

Hence we expect that the TCMOT could also be useful for cadmium \cite{PhysRevA.76.043411}, dysprosium \cite{PhysRevA.82.043425,PhysRevLett.107.190401}, erbium \cite{PhysRevA.85.051401}, and thulium \cite{PhysRevLett.96.143005,QuantumElectron.44.515}. Cadmium trapping could especially benefit from the TCMOT, as in Ref. \cite{PhysRevA.76.043411}, $B'=500$ G/cm was used for the broad linewidth MOT. We note, however, that for cadmium, trapping by the narrow transition has not yet been observed, likely due to two photon ionization by the MOT light.

\section{Conclusion}
We have observed a TCMOT for ytterbium atoms at magnetic field gradients as low as $B'=2$ G/cm by applying light on the singlet and triplet transitions simultaneously, yielding an atom number similar to a conventional singlet MOT at $15\leq B' \leq 45$ G/cm. The TCMOT trapped $\sim10^5$ atoms and attained a density up to $10^9$ cm$^{-3}$ at a temperature of 1 mK. The atom number can be increased by means of a Zeeman slower, while the final temperature can be lowered by cooling only with 556 nm light at reduced intensity after collecting the atoms \cite{PhysRevA.87.013615}.  The TCMOT could be beneficial in situations where a large magnetic field gradient is not permitted or desired.  

\begin{acknowledgements}
This research was supported by NSF and DARPA QuASAR. B.B. acknowledges support from the National Science and Engineering Research Council of Canada and Q.Y. from the Undergraduate Research Opportunity Program at MIT. We would like to thank Wolfgang Ketterle for insightful discussions.
\end{acknowledgements}

\bibliographystyle{apsrev4-1}
\bibliography{AMOTechnique,AtomicClock,QuantumGas,QuantumMeasurement}

\end{document}